\documentclass[twocolumn,pra,longbibliography,nobalancelastpage,aps,raggedbottom, superscriptaddress]{revtex4-2}
\usepackage{amsmath,amsfonts,amssymb,scalerel,natbib}
\usepackage{graphicx,hyperref,epstopdf,xcolor,tikz}
\usepackage{array,mathptmx,textcomp,braket,times}
\usepackage{tabularx,makecell,multirow}
\usepackage[utf8]{inputenc}
\usepackage[T1]{fontenc}

\definecolor{lime}{HTML}{A6CE39}
\DeclareRobustCommand{\orcidicon}
{
	\begin{tikzpicture} 
	\draw[lime, fill=lime] (0,0) circle [radius=0.15] node[white] {{\fontfamily{qag}\selectfont \tiny ID}};
	\draw[white, fill=white] (-0.0625,0.095) 	circle [radius=0.007];
	\end{tikzpicture}
	\hspace{-2.2mm}
}
\newcommand\orcidID[1]{\href{https://orcid.org/#1}{\orcidicon}}

\newcommand{\be}{\begin {equation}}
\newcommand{\ee}{\end {equation}}
\newcommand{\beqa}{\begin {eqnarray}}
\newcommand{\eeqa}{\end {eqnarray}}
\newcommand{\mb}{\mathbf}
\newcommand{\Sch}{Schr\"odinger }

\hypersetup{colorlinks,citecolor=blue,filecolor=black,linkcolor=blue,urlcolor=blue}

\begin{document}

\title{Sub-cycle pulse control of holographic and non-holographic electron interferences}

\author{Rambabu Rajpoot\orcidID{0000-0002-2196-6133}}
\email[E-mail: ]{rambabu.rajpoot@riken.jp}
\affiliation{Extreme Photonics Research Team, RIKEN Center for Advanced Photonics, RIKEN, 2-1 Hirosawa, Wako, Saitama 351-0198, Japan}

\author{Eiji J. Takahashi\orcidID{0000-0003-1672-1171}}
\email[Corresponding author: ]{ejtak@riken.jp}
\affiliation{Extreme Photonics Research Team, RIKEN Center for Advanced Photonics, RIKEN, 2-1 Hirosawa, Wako, Saitama 351-0198, Japan}

\date{\today}

\begin{abstract}

We investigate the influence of sub-cycle laser pulses on holographic and non-holographic intracycle interferences by analyzing the photoelectron momentum distributions of helium using TDSE simulations supported by classical trajectory calculations. The results show that the forward-scattering holographic (FSH), backward-scattering holographic (BSH), and time double-slit (TDS) structures are found to be highly sensitive to the pulse duration, carrier-envelope phase (CEP), and temporal envelope in the sub-cycle regime. Sub-cycle pulses with CEP values of $0^\circ$ and $90^\circ$ selectively enhance or suppress distinct features, isolating holographic patterns and enhancing BSH fringes. Classical analysis reveals that the intrinsic chirp inherent to sub-cycle fields shortens the recollision time for scattering trajectories, thereby increasing the fringe spacing in FSH and BSH patterns, while simultaneously enlarging the ATI peak spacing associated with TDS interference. Pulse envelope variations, even at fixed FWHM duration, further reshape the fringe spacings by modifying the instantaneous frequency and vector potential slope near ionization times. These results demonstrate that sub-cycle pulses enable precise temporal control of holographic interference, offering new opportunities for probing and manipulating attosecond electron dynamics.
\end{abstract}

\maketitle

\section{Introduction}
\label{sec1}

Strong-field induced tunnel ionization is one of the central processes governing light-matter interactions in atoms and molecules. Once ionized, the electron undergoes rapid sub-cycle acceleration in the laser field, giving rise to a wide range of ultrafast phenomena, including high-order harmonic generation (HHG)~\cite{Ferray1988_JPhysB, Krause1992_PRL, Agostini2004_RepProgPhys, Corkum2007_NatPhy, Krausz2009_RMP}, high-order above-threshold ionization (HATI)~\cite{Schafer1993_PRL, Paulus1994_PRL, Becker2002_AdvAtMolPhys}, and non-sequential double ionization (NSDI)~\cite{Fittinghoff1992_PRL, Walker1994_PRL, Becker2012_RevModPhys}. Of particular interest are those electrons that experience rescattering with the parent ion. Their sub-angstrom spatial resolution and attosecond temporal sensitivity form the basis of strong-field photoelectron holography (SPH)~\cite{Barton1988_PRL, Huismans2011_Science, Bian2011_PRA, Huismans2012_PhysRevLett, Haertelt2016_PRL, Shvetsov2018_PRA, Faria2020_RepProgPhys}, which has emerged as a powerful tool for probing ultrafast structural and dynamical information in atomic and molecular systems.

The photoelectron momentum distribution (PMD) encodes these electron dynamics through a variety of interferometric features~\cite{Remetter2006_NatPhys}. Depending on whether the interfering wave packets originate within the same optical cycle or in different cycles, the resulting patterns are classified as intracycle or intercycle interferences. Intercycle interference produces the well-known concentric above-threshold ionization (ATI) rings~\cite{Schafer1990_PRA, Schafer1993_PRL}, whereas intracycle interference gives rise to a rich set of holographic structures, including the spider-leg-like forward-scattering (FSH) and fish-bone-like backscattering (BSH) patterns~\cite{Bian2011_PRA}. Under suitable field conditions, a non-holographic intracycle interference, namely the time double-slit (TDS) interference fringes, emerges from two non-scattering electron trajectories ionized within a single laser cycle~\cite{Gopal2009_PRL}.

However, in multi-cycle laser fields, the holographic structures of interest are often obscured by strong intercycle contributions. The use of near-single-cycle pulses has therefore proven essential for isolating sub-cycle electron dynamics and enhancing the visibility of holographic features~\cite{Du2016_OptLett, Murakami2020_PRA, Taoutioui2022_SciRep, Khurelbaatar2024_LightSci}. Rapid advances in ultrafast laser technology now enable the generation of pulses with full-width at half-maximum (FWHM) durations shorter than one optical cycle~\cite{Nishimiya2025_AdvPhysX, Xu2024_NatPhoton, Nishimiya2024_OptLett, Rossi2020, Nie2018_NatPhoton, Liang2017, Chu_2016, Balciunas2015_NatCommun, Voronin2013_OptLett, Krauss2010_NatPhoton}. Such sub-cycle pulses not only suppress intercycle interference but also allow direct control over the appearance of FSH, BSH, and time double-slit patterns through carrier-envelope phase (CEP) variation~\cite{Yuan2021_PRA, Murakami2020_PRA}.

A key feature of sub-cycle fields is the presence of an \emph{intrinsic chirp} arising from the temporal confinement of the pulse. This chirp manifests as a temporal phase analogous to the Gouy phase of spatially focused beams and results in a blueshift of the instantaneous frequency~\cite{LinSubcycle-2006}. The intrinsic chirp has been experimentally observed in sub-cycle terahertz pulses~\cite{Lin2010_PRA} and plays an increasingly recognized role in shaping strong-field ionization dynamics~\cite{LinSubcycle-2006, Cai2016_OptCommun}. In addition, the temporal envelope of the pulse can significantly reshape the driving field and thereby influence the phase accumulation between interfering electron trajectories. Indeed, sub-cycle pulse duration and envelope shape have already been shown to strongly affect HHG emission and related strong-field processes~\cite{Rajpoot2025_PRR, Zheng_2011, Holkundkar2023_PLA, Nishimiya2025_arXiv}.

Because holographic structures arise from the coherent interference of electron trajectories with distinct ionization and return times, they are highly sensitive to these temporal distortions. Accounting for the combined effects of intrinsic chirp and envelope-induced waveform reshaping is therefore essential for understanding and manipulating holographic interference in the sub-cycle regime.

In this work, we investigate the PMDs of helium driven by sub-cycle laser pulses, focusing specifically on how the pulse cycle number (and associated intrinsic chirp) and the envelope function modify the FSH, BSH, and TDS intracycle interference structures.

The paper is organized as follows. Section~\ref{sec2} describes the theoretical methods and the sub-cycle field profiles employed. Section~\ref{sec3} presents and analyzes the resulting interference structures. Section~\ref{sec4} summarizes our findings and provides concluding remarks. Atomic units (a.u.) are used throughout unless otherwise stated.

\section{Theoretical Methods}
\label{sec2} 
In the following, we present the simulation details of the PMD calculation, followed by a discussion on the sub-cycle field profiles used in this work.

\subsection{Simulation details for TDSE}
\label{sec2_tdse}
To simulate the PMD, we numerically solve the three dimensional (3D) time-dependent \Sch equation (TDSE) within the single-active-electron (SAE) approximation. The TDSE describing the laser-atom interaction under the dipole approximation is expressed as
\be
	i \partial_t \ket{\psi(t)} = \left[- \frac{1}{2}\nabla^2 + V(r) + H_{int} -i V_{abs}(r) \right] \ket{\psi(t)} ,
 \ee
where $H_{int} = -i\mb{A}(t)\cdot\nabla$ represents the electron-field interaction in the velocity gauge. The vector potential is related to the electric field $\mb{E}(t)$ through $\mb{A}(t) = -\int_0^t \mb{E}(t') dt'$. $V_{abs}(r)$ is an absorbing potential introduced to suppress spurious reflections of the wave function from the numerical boundary. The atomic binding potential $V(r)$ is defined as
\be 
V(r) =
\begin{cases}
	-(1 + e^{-a_0 r})/r, &  r<R_{co},\\
	-(2R_{co} - r)/R_{co}^2, & R_{co} < r < 2R_{co},\\
	0, & r > 2R_{co}.
\end{cases}
\ee
Beyond the cutoff radius $R_{co}$, the effective Coulomb potential transitions to a linear form and vanishes for $r>2R_{co}$. The parameter $a_0 = 1.91$ is chosen such that the ground-state ionization potential of helium matches the experimental value, $I_p = 0.90$ a.u. The initial ground state is obtained through imaginary-time propagation.

The PMD amplitude $a(\mb{k})$ at time $T_f$ after the laser pulse can be approximated as
\begin{multline} 
a(\mb{k}) = \int_0^{T_f} dt \partial_t \braket{\mb{k}(t)|\Theta(r-R_l)|\psi(t)}\\+ \braket{\mb{k}(0)|\Theta(r-R_l)|\psi(0)}. 
\end{multline}
where,  $\mb{k}(t)$ denotes the Volkov wave function, $\Theta(r-R_l)$ is the Heaviside step function, and $R_l$ is a sufficiently large radius beyond which the initial wave function $\psi(0)$ becomes negligible. The PMD calculations are performed by numerically solving the 3D-TDSE within the SAE framework using the Qprop package~\cite{Tulsky2020_ComputPhysCommun}.

In our simulations, the radial simulation domain of $1000$ a.u. is considered with the cutoff radius $R_{co} = 300 $ a.u., and the flux capturing surface is positioned at $R_l = 800$ a.u. The number of partial waves is limited to $L_{max} = 100$. We use the grid with a radial interval of $0.2$ a.u. and a time step of $0.05$ a.u., to ensure convergence of the calculations.

\subsection{Classical-trajectory based model}
\label{sec2_classic}
The holographic structures observed in PMDs originate from the interference between electrons that reach the detector the with same final momentum. Among these, one electron undergoes rescattering off the parent ion before reaching the detector, while the other propagates directly without rescattering. In this subsection, we employ a simplified description of electron dynamics in a laser field based on the well-known three-step recollision model ~\cite{Schafer1993_PRL, Corkum1993_PRL, Bandrauk20012_JModOpt, Bian2011_PRA}. 

The time-dependent velocity and position of a electron ionized at time $t_i$ can be obtained as
\begin{align} 
 \mb{v}(t) &= -\int_{t_i}^t \mb{E}(t') dt' + \mb{v}_i,
 \label{velocity}\\
 \mb{r}(t) &= \int_{t_i}^t \mb{v}(t') dt' + \mb{r}_i,
 \label{position}
\end{align}
where, $\mb{r}_i$, and $\mb{v}_i$ are the initial position, and velocity, respectively. For the linearly polarized field $\mb{E}(t)$ described by Eq.~(\ref{laser}), the electron is assumed to be released along the polarization direction of the field at a distance $|z_i| = I_p/E_0$ from the ionic core, i.e., $r_{\parallel}^i = \pm |z_i|$ ($r_{\parallel}^i$ has sign opposite to $\mb{E}(t_i)$ at ionization time $t_i$), and $r_{\perp}^i = 0$. The initial velocity of rescattering electron is $\mb{v}_i = 0$, whereas for the direct electron, $v_{\parallel}^i = 0$, and $ v_{\perp}^i \neq 0$. In Eqs.~(\ref{velocity}) and (\ref{position}), the influence of the Coulomb potential is neglected due to the large initial distance $z_i$ and excursion amplitude $r_{\alpha} = E_0/\omega_0^2$ for the laser parameters used in this work. 

The phase of the electron along any trajectory is given by the classical action~\cite{Yuan2021_PRA}: 
\be
 S = \int_{t_i}^{T_f} \left[ \frac{1}{2} \mb{v}^2(t) + I_p \right] dt, 
\ee
where, $T_f$ denotes the pulse turn-off time. The interference map in the momentum plane ($p_{\parallel}, p_{\perp}$) is calculated by $\cos^2(\Delta S/2)$, where $\Delta S$ is the phase difference between the direct and rescattering electron trajectories. We have also weighted each tunneling trajectory by the instantaneous tunneling ionization probability calculated using the Ammosov-Delone-Krainov (ADK) formula~\cite{ADK1986_JETP}. 

\subsection{Sub-cycle field profile}
\label{sec2_laser}
In this work, the electric field of the linearly polarized pulse with an arbitrary pulse duration is described using the sub-cycle pulsed beam (SCPB) model~\cite{LinSubcycle-2006}. The SCPBs are derived from the oscillating dipole model through the complex-source point method \cite{PhysRevE.67.016503,Heyman-89}. The field expressions so obtained are exact solutions of Maxwell's equations. In the dipole approximation and plane-wave limit, the electric field of a linearly polarized laser pulse is given as,
\be
\mb{E}(t) = \text{Re}\left\{ \frac{A_d(t')}{|A_d(0)|} f(t')\ E_0\exp(i\omega_0t' + i\phi_0) \right\} \hat{\mb{e}}_z,
\label{laser}
\ee
where $E_0$ is the peak field amplitude, $\omega_0$ is the carrier frequency, $\phi_0$ represents the carrier-envelope phase (CEP), and $t' = t - t_0 - x/c$ is the retarded time (with $x = 0$ and $t_0$ denoting the pulse center). The complex time-dependent function $A_d(t')$ is defined as~\cite{Zheng_2011}:
\be
A_d(t') = 1 - \ddot{f}/[\omega_0^2 f(t')] - 2i \dot{f}/[\omega_0 f(t')],
\ee 
where $\dot{f}$ and $\ddot{f}$ are the first- and second-order derivatives of the envelope function $f(t')$ with respect to the retarded time $t'$. The total phase of the pulse is
\be \phi(t') = \omega_0 t' + \phi_0 + \text{arg}[A_d(t')], 
\label{total_phase}
\ee 
and the corresponding instantaneous frequency of the pulse can be obtained by taking the derivative of the total phase as $\omega(t') = d\phi(t')/dt'$. In this work, we utilized the following three envelope functions to model the sub-cycle field profiles~\cite{Zheng_2011}:
\be 
f(t') =
\begin{cases}
	\exp[-2\text{ln}2 (t'/\tau)^2], &  \text{Gaussian},\\
	\text{sech}(1.76 t'/\tau), &  \text{Hyperbolic secant},\\
	1/[1 + (1.29 t'/\tau)^2], &  \text{Lorentzian},
\end{cases}
\label{laser_envlp}
\ee
where, $\tau$ is the full width at half maximum (FWHM) of the intensity envelope function $|f(t')|^2$. For all the numerical simulations presented in this work, we employ a linearly polarized pulse with wavelength $\lambda_0 = 800$ nm (corresponding to a frequency $\omega_0 = 0.057$ a.u.) and peak intensity $I_0 = 5\times 10^{14}$ W$/$cm$^{2}$. These parameters ensure that the ionization of helium occurs predominantly via tunneling, as indicated by the Keldysh parameter $\gamma = \omega_0 \sqrt{2I_p}/E_0 < 1$, where $E_0$ is the peak field amplitude in a.u.~\cite{Keldysh1965_JETP}.
\begin{figure}[t]
\centering\includegraphics[width=1\columnwidth]{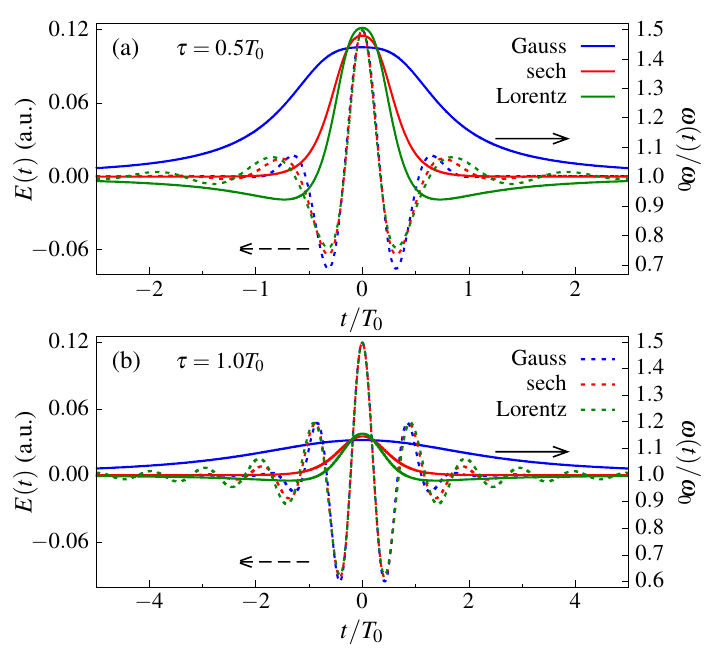}
\caption{Temporal profiles of the laser electric field $\mb{E}(t)$ (dashed curve, left-axis) and instantaneous frequency $\omega(t)$ (solid curve, right-axis) for different envelope functions defined in Eq.~\ref{laser_envlp} for FWHM durations: (a) $\tau = 0.5T_0$, and (b) $\tau = 1.0T_0$. The laser pulses are of $800$ nm wavelength with peak intensity $I_0 = 5\times 10^{14}$ W$/$cm$^{2}$, and CEP $\phi_0 = 0^\circ$. Here, $T_0\ (= 2\pi/\omega_0)$ denotes one laser cycle. }
\label{fig_laser_wt}
\end{figure}

Figure~\ref{fig_laser_wt} compares the electric fields $\mb{E}(t)$ and the corresponding instantaneous frequencies $\omega(t)$ of half-cycle and single-cycle FWHM pulses for different envelope functions defined in Eq.~\ref{laser_envlp}. The time-dependence of the laser frequency introduces an intrinsic chirp in the pulse~\cite{Lin2010_PRA}, which becomes increasingly pronounced as the pulse duration shortens from a single-cycle to the sub-cycle regime. This intrinsic chirp manifests as a temporal compression or ``shrinking'' of the electric field waveform in sub-cycle pulses. Notably, the extent of this modulation depends on the choice of the envelope function. Given that the waveform of the laser filed plays crucial role in the electron dynamics and that these dynamics are directly imprinted in the PMD, it is essential to account for the intrinsic chirp when investigating PMDs generated by such short-duration pulses. Consequently, it is of particular interest to examine how the interference structures in PMD evolves as the pulse duration decreases from a single-cycle to the sub-cycle regime.

\section{Results and Discussion}
\label{sec3}
In this section, we first discuss the salient features of the PMDs obtained using single- and half-cycle laser pulses. We identify the primary interference contribution, including FSH, BSH, and TDS intracycle interference patterns. Subsequently, we investigate how these interference structures evolve as the pulse duration enters the sub-cycle regime, and finally examine the role of the pulse envelope in shaping the resulting interference features.

\subsection{PMD characteristics for single- and half-cycle pulses}
\label{subsec1}

\begin{figure}[t]
\centering\includegraphics[width=1\columnwidth]{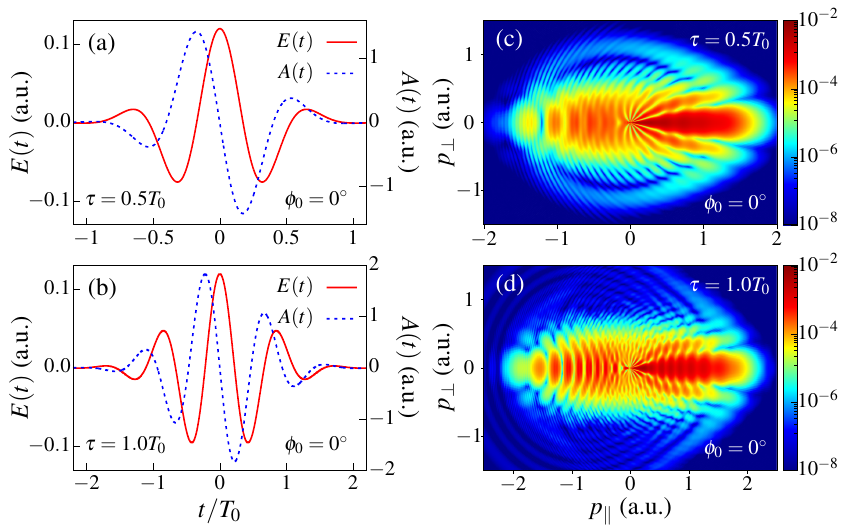}
\caption{ Half-cycle and single-cycle laser fields with CEP $\phi_0 = 0^\circ$ and corresponding TDSE-simulated PMDs of helium. The temporal profiles of the laser electric field $\mb{E}(t)$ (red solid curve, left-axis) and vector potential $\mb{A}(t)$ (blue dashed curve, right-axis) for Gaussian-enveloped pulses with FWHM $\tau:$ (a) $0.5T_0$ and (b) $1.0T_0$. The corresponding PMDs are shown in (c) and (d), respectively. } 
\label{fig2}
\end{figure}

Figure~\ref{fig2} presents the laser waveforms and the corresponding TDSE-simulated PMDs for half-cycle and single-cycle Gaussian pulses with CEP $\phi_0 = 0^\circ$. As expected, the electric field is symmetric about the pulse maximum, while the vector potential exhibits antisymmetric temporal behavior. For an electron released at time $t_i$ with negligible initial velocity and subsequently driven only by the laser field, the final momentum is given by $\mb{p}_f = -\mb{A}(t_i)$.

The PMDs in Fig.~\ref{fig2} display the characteristic interference structures induced by linearly polarized fields~\cite{Bian2011_PRA, Marchenko2011_PRA}. The momentum component $p_{\parallel}$ is along the laser polarization direction. Since ionization is most probable after the field maximum, electrons predominantly populate the $p_{\parallel} > 0$ region, corresponding to the negative vector potential in the trailing half-cycle [since $\mb{p}_f = -\mb{A}(t_i)$]. The short pulse duration suppresses intercycle interference, and thus the familiar above-threshold ionization (ATI) rings are absent ~\cite{Telnov2009_PRA}.

For the half-cycle driver [Fig.~\ref{fig2}(c)], both the FSH and BSH features are clearly resolved. The FSH structure appears as a spider-lig-like pattern in the $p_{\parallel} > 0$ region, whereas the BSH pattern emerges as right-curved fringes in the $p_{\parallel} < 0$ region. Notably, the half-cycle pulses isolates the BSH interference, which is often obscured in longer pulses by overlapping contributions from intracycle and FSH pathways. Since backscattering electrons have smaller impact parameters, the BSH pattern encodes enhanced structural information, making its clear visibility particularly valuable for strong-field holography. 

For the single-cycle pulses [Fig.~\ref{fig2}(d)], the PMD exhibits increased interference complexity. The spider-leg-like FSH pattern remains predominant for $p_{\parallel}>0$, but now coexists with intracycle interference fringes (right-curved stripes). On the $p_{\parallel}<0$ side, the BSH contributions persists but are largely buried under stronger intracycle interferences (left-curved stripes), illustrating the increased complexity of interference pathways as the pulse duration increases.

\begin{figure}[t]
\centering\includegraphics[width=1\columnwidth]{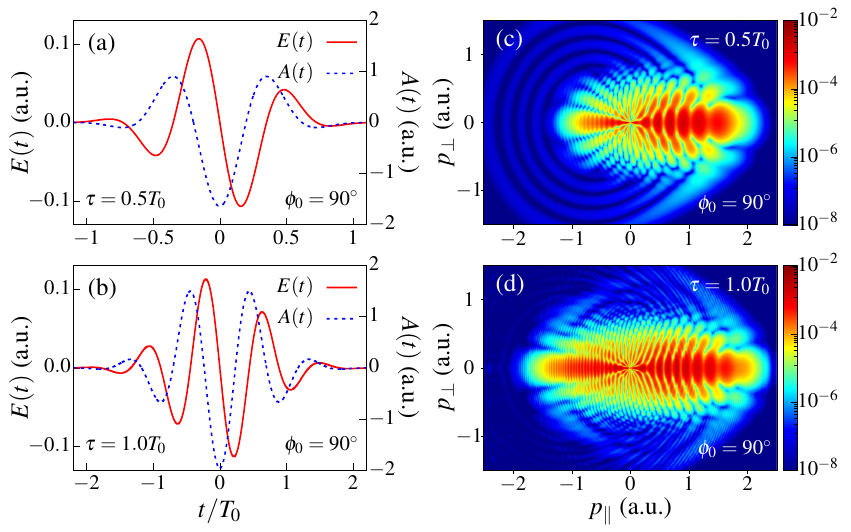}
\caption{ Similar to Fig. \ref{fig2}, but with CEP $\phi_0 = 90^\circ$. (a) and (b) shows the temporal profiles of the electric field $\mb{E}(t)$ and its corresponding vector potential $\mb{A}(t)$ for laser pulses with FWHM $\tau = 0.5T_0$ and $1.0T_0$, respectively. The corresponding PMDs of helium atom are shown in (c) and (d), respectively. } 
\label{fig3}
\end{figure}

Next, we analyze PMDs produced by half-cycle and single-cycle pulses with CEP $\phi_0 = 90^\circ$, shown in Fig.~\ref{fig3}. Unlike the $\phi_0 = 0^\circ$ case, here the electric field is temporally antisymmetric, whereas the vector potential is symmetric about the pulse maximum. This waveform is particularly favorable for the emergence of time double-slit interference~\cite{Gopal2009_PRL}, characterized by fringe spacings that increase with photoelectron energy~\cite{Diego2006_PRA, Diego2010_PRA}. This intracycle interference arise when two electron trajectories, born at times $t_{i1}$ and $t_{i2}$ with $A(t_{i1}) = A(t_{i2})$ and opposite electric field signs $E(t_{i1}) = -E(t_{i1})$, interfere. The resulting PMDs exhibit a spider-leg-like pattern in the $p_{\parallel}<0$ region and a fish-bone-like structure in the $p_{\parallel}>0$ region [Figs.~\ref{fig3}(c) and~\ref{fig3}(d)]. The negative parallel-momentum side spider-leg-like pattern originates from electrons released in the quarter-cycle following the negative field peak after the pulse center, leading to the FSH structure with comparatively low electron yield. The reduced yield occurs because the subsequent positive half-cycle of the electric field is insufficient to reverse the direction of all liberated electrons. As the pulse duration increases from the half-cycle to single-cycle, the asymmetry between positive and negative parallel-momentum side yields diminishes.

The fish-bone-like interference observed on the $p_{\parallel}>0$ side arises from overlapping contributions of BSH and time double-slit pathways. Because the fringe positions and densities produced by these two mechanisms nearly coincide, disentangling them is challenging. Owing to the relatively small rescattering cross section, the time double-slit interference dominates the PMD. Moreover, the stripe (fringe) density is greater for the single-cycle pulse than for the half-cycle case. This behavior originates from the intrinsic chirp and envelop-induced field modulation of the pulse~\cite{Taoutioui2022_SciRep}. For longer pulses, electron experience higher and more sustained electric fields, resulting in a large phase accumulation and consequently a denser fringe pattern. These observations naturally motivate a closer examination of how the interference structures evolve as the pulse duration enters the sub-cycle regime.

\subsection{Effect of pulse length}
\label{subsec2}
\begin{figure}[t]
\centering\includegraphics[width=1\columnwidth]{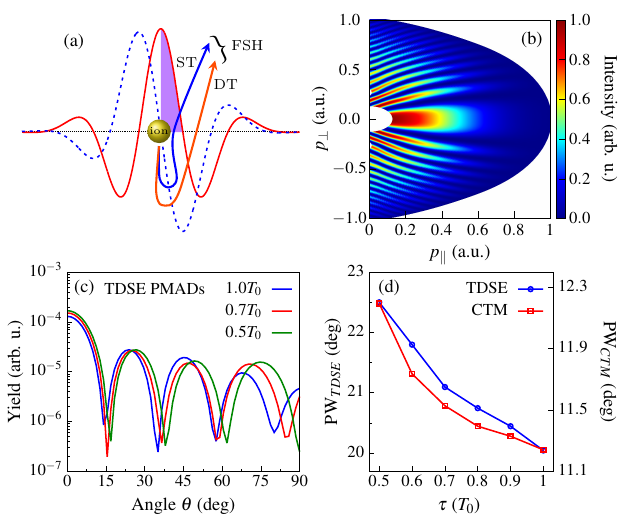}
\caption{Variation in stripe density of the FSH structure. (a) Schematic illustration of underlying mechanism contributing to the observed FSH structure. The half-cycle laser field $\mb{E}(t)$ (red solid curve) with CEP $\phi_0 = 0^\circ$ and the corresponding vector potential $\mb{A}(t)$ (blue dashed curve) are shown. The shaded quarter-cycle indicates the ionization times of the direct and forward-scattering trajectories. (b) PMDs showing the spider-leg-like FSH interference pattern obtained from the classical-trajectory based model (CTM). (c) PMADs for different pulse durations, extracted from the TDSE-simulated PMDs of helium. (d) Average peak widths (PWs) of PMADs-calculated using TDSE and CTM methods for different pulse durations. }
\label{fig4}
\end{figure}

\begin{figure}[t]
\centering\includegraphics[width=1\columnwidth]{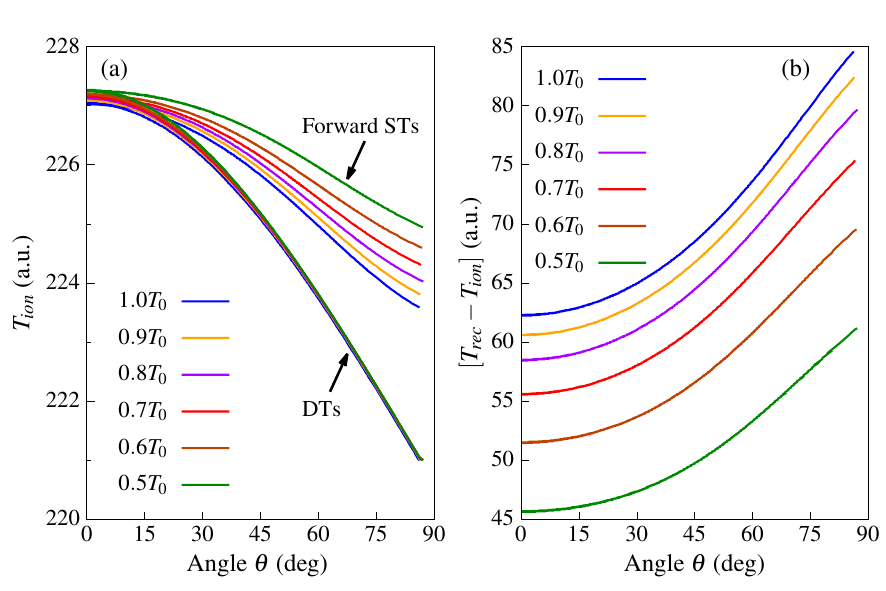}
\caption{(a) Classical ionization times $T_{ion}$ of the direct (DT) and scattering trajectories (STs) as a function of the polar angle $\theta$. (b) The recollion time $\Delta T$ of  STs, defined as the difference between the ionization time $T_{ion}$ and the rescattering time $T_{rec}$ for different pulse durations. All data points are extracted at a fixed final momentum $p = 0.5$ a.u. from the classical model based PMDs. }
\label{fig5}
\end{figure}

So far, we have established that the holographic structures are sensitive to the duration of driving laser pulse, exhibiting a variation in stripe (fringe) density with pulse duration. We now analyze in detail how the fringe spacing varies with the cycle number of the driving pulse. 

We first consider the case of the FSH structure, observed for CEP $\phi_0 = 0^\circ$ pulses [Figs.~\ref{fig2}(c,d)]. Figure~\ref{fig4} illustrates how the pulse duration influences the stripe density of the FSH pattern. In Fig.~\ref{fig4}(a), a schematic diagram depicts the physical origin of the FSH interference. The red solid curve represents the electric field $\mb{E}(t)$ of a half-cycle (FWHM) pulse, while the blue dashed curve shows the corresponding vector potential $\mb{A}(t)$. The shaded quarter-cycle indicates the ionization times of the direct (DT) and scattering (ST) trajectories that contribute to the observed interference. The electrons are initially ionized in the $p_{\parallel}<0$ direction [opposite to the instantaneous field $\mb{E}(t_i)$] with different transverse momenta. The electrons are then redirected by the laser field when the electric field reverses sign. One subset of electrons reaches the detector directly (direct electrons), while another subset is forward-scattered toward the detector (scattered electrons). The coherent interference between these two classes of electron trajectories gives rise to the characteristic spider-leg-like FSH pattern. Figure~\ref{fig4}(b) presents the interference pattern generated by direct and forward-scattering trajectories ionized during the shaded quarter-cycle of laser field. The PMDs are obtained using the classical-trajectory based model (CTM) [described in subsection~\ref{sec2_classic}]. This simplified model neglects the Coulomb potential and therefore cannot quantitatively reproduce the TDSE results. However, there is a clear spider-leg-like pattern for $p_{\parallel}>0$, which is the characteristic spectral feature of FSH interference and is in qualitative agreement with the TDSE-simulated PMDs shown in Fig.~\ref{fig2}(c).

To analyze the density of the spider-legs (FSH interference fringes), we extract the photoelectron momentum angular distributions (PMADs) by integrating the PMDs over a narrow momentum range. Figure~\ref{fig4}(c) presents the PMADs obtained from the TDSE-simulated PMDs within the momentum interval $0.01 < p < 0.1$ a.u. for different pulse durations. Owing to the vertical symmetry of the PMDs along the $p_{\parallel}$-axis, the PMADs are evaluated as a function of the polar angle $\theta$ [$= \tan^{-1}(p_{\perp}/p_{\parallel})$] in the range $0^\circ < \theta < 90^\circ$. The peaks of the PMADs are the most densely spaced for the $\tau = 1.0T_0$ pulse and become progressively sparser as the pulse duration decreases.

Furthermore, the quantitative dependence of the FSH stripe density on the pulse duration is shown in Fig.~\ref{fig4}(d). The average peak widths (PWs) of the PMADs are calculated using the TDSE (blue curve, left-axis) and the CTM (red curve, right-axis) methods. In both methods, the average PW is largest for $\tau=0.5T_0$ pulse and  decreases as the pulse duration increases to $\tau=1.0T_0$. A larger PW corresponds to a lower stripe density in the observed FSH pattern. Although the number of FSH stripes in the classically calculated PMDs [Fig.~\ref{fig4}(b)] differs from those in the TDSE results [Fig.~\ref{fig2}(c)], the overall trend—namely, a reduction of stripe density for shorter pulses—is consistent across both the methods, as highlighted in Fig.~\ref{fig4}(d). In the classical model, the PMADs originate from the interference of electrons ionized within the quarter-cycle immediately following the pulse maximum [see Fig.~\ref{fig4}(a)]. While the peak intensity remains unchanged for pulses of different durations, the timing at which the electric field attains a specific value varies because of the difference in the intrinsic chirp of the pulses. We therefore infer that the feature of PMADs is closely related to the temporal characteristics of the underlying electron trajectories. To substantiate this connection, we analyzed the ionization and scattering times of different trajectories, following the approach discussed in Ref.~\cite{Yuan2021_PRA}. 

Figure~\ref{fig5}(a) shows the classical ionization times of the direct and forward scattering trajectories as a function of the polar angle $\theta$. For clarity, the ionization times are shown for trajectories with a fixed final momentum of $p = 0.5$ a.u. The ionization times $T_{ion}$ of the direct trajectories (DTs) are similar for all the pulse durations, whereas those of the scattering trajectories (STs) exhibit a noticeable delay with increasing $\theta$. It should be noted that an increase in $\theta [= \tan^{-1}(p_{\perp}/p_{\parallel})]$ corresponds to an increase in the final perpendicular momentum $p_{\perp}$ within the considered range $0^\circ < \theta < 90^\circ$. Since the ionization times of the DTs are nearly identical across pulse durations at a fixed polar angle, they do not  contribute to the relative shift of the PMAD peaks with changing pulse duration~\cite{Yuan2021_PRA}. In contrast, the ionization times of the STs vary with pulse duration. 

Figure~\ref{fig5}(b) shows the recollision time ($\Delta T$), defined as the difference between the ionization time ($T_{ion}$) and the rescattering time ($T_{rec}$) for different pulse durations. The recollision time $\Delta T$ increases with increasing $\theta$ (hence with $p_{\perp}$), but decreases as the pulse duration shortens from $1.0T_0$ to $0.5T_0$. Moreover, the relative shift among the recollision-time curves grows stronger as the pulse duration decreases, resembling the behavior of the intrinsic chirp, which becomes more pronounced for shorter pulses. Therefore, it can be concluded that the stripe density in the FSH structure is primarily governed by the recollision time: the larger the time difference between ionization and rescattering, the higher the stripe density. This observation is consistent with the findings of Refs.~\cite{Yuan2021_PRA, Huismans2012_PhysRevLett, Murakami2020_PRA}, where the influence of laser wavelength and CEP on stripe density was investigated. Consequently, we infer that the stripe density in the FSH structure is strongly influenced by the intrinsic chirp of sub-cycle laser pulses. 

\begin{figure}[t]
\centering\includegraphics[width=1\columnwidth]{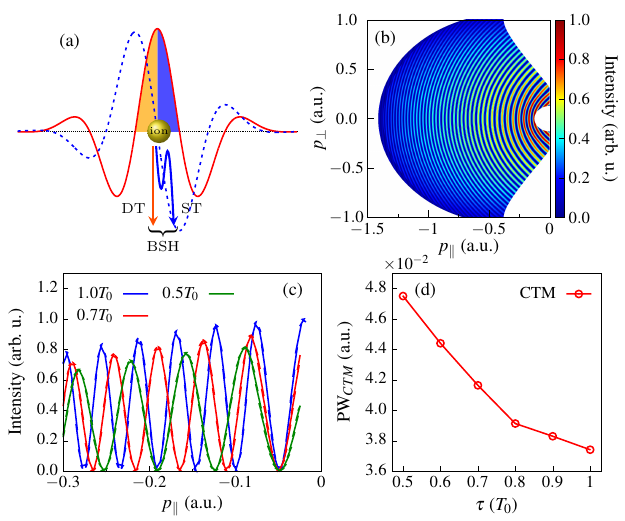}
\caption{Variation in fringe spacing of the BSH structure. (a) Schematic illustration of the mechanism responsible for the observed BSH pattern. The half-cycle laser field $\mb{E}(t)$ (red solid curve) with CEP $\phi_0 = 0^\circ$ and the associated vector potential $\mb{A}(t)$ (blue dashed curve) are also depicted. The quarter-cycles shaded in orange and blue represent the ionization times of the direct and backscattering trajectories, respectively. (b) PMDs showing the BSH interference pattern obtained from the classical-trajectory based model. (c) Line profiles of the BSH structure at $p_{\perp} = 0.0$ a.u. for different pulse durations, calculated using the classical method. (d) Corresponding average peak widths of BSH fringes for different pulse durations.}
\label{fig6}
\end{figure}

Next, we examine the BSH structure that appears in the negative parallel-momentum region for CEP $\phi_0 = 0^\circ$ pulses [Fig.~\ref{fig2}(c)]. Figure~\ref{fig6} shows how the pulse duration affects the spacing of the BSH interference fringes. In Fig.~\ref{fig6}(a), the quarter-cycles shaded in orange and blue indicate the ionization times of the direct and scattering trajectories, respectively, that contribute to the observed interference. The DTs are born in the quarter-cycle where both the electric field $\mb{E}(t)$ and the vector potential $\mb{A}(t)$ have positive amplitudes and directly reach the detector. In contrast, the STs originate in the subsequent quarter-cycle, where the electric field maintains its sign but the vector potential reverses. These electrons are driven back toward the core, undergo backward scattering, and coherently interfere with the DTs, forming the BSH structure. Backward-scattered electrons approach the ionic core more closely, making them particularly sensitive to the target structure and useful for dynamic imaging~\cite{Bian2012_PRL}. Figure~\ref{fig6}(b) shows the BSH fringes obtained from the classical-trajectory model, which qualitatively reproduce the rightward-curved interference pattern observed in the TDSE-simulated PMDs [Fig.~\ref{fig2}(c)] for half-cycle driving fields. The BSH interference is clearly visible for half-cycle pulses but becomes increasingly obscured by intracycle interferences as the pulse duration increases. Consequently, extracting a distinct BSH line profile from the TDSE-simulated PMDs becomes impractical. We, therefore rely on the classical-trajectory model to investigate how the pulse duration influences the BSH fringe characteristics. 

\begin{figure}[t]
\centering\includegraphics[width=1\columnwidth]{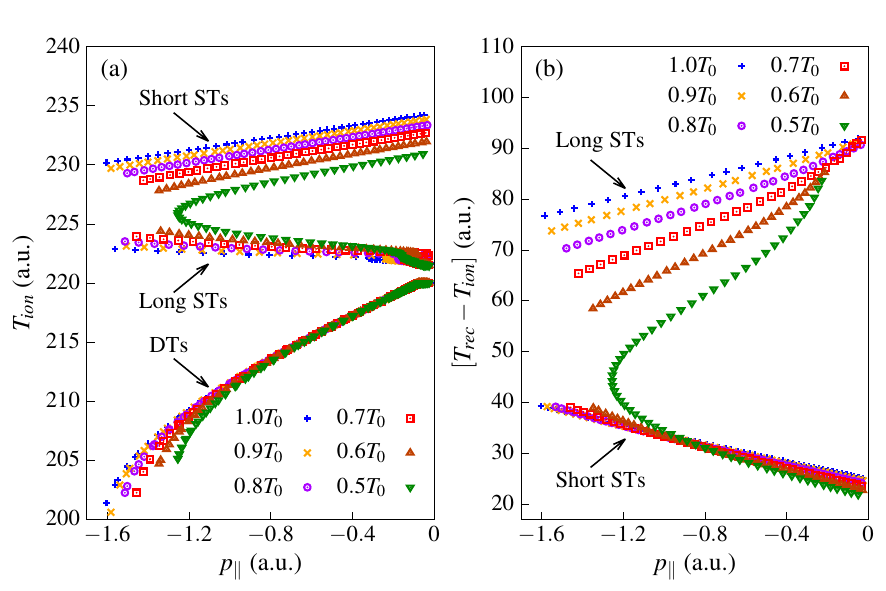}
\caption{(a) Classical ionization times $T_{ion}$ of the direct (DT) and scattering trajectories (STs) as a function of the final parallel momentum $p_{\parallel}$. (b) The recollion time $\Delta T$ of  STs, defined as the difference between the ionization time $T_{ion}$ and the rescattering time $T_{rec}$ for different pulse durations. All data points are extracted at the zero final perpendicular-momentum $p_{\perp}$ from the classical model based PMDs. }
\label{fig7}
\end{figure}

Figure~\ref{fig6}(c) shows the line profiles of the BSH structure at $p_{\perp} = 0$ a.u. for different pulse durations, obtained from the classical model. The effect of the frequency blueshift is clearly reflected in the relative spacing of the BSH interference peaks. The fringes are most densely packed for $\tau = 1.0T_0$ and gradually become more widely spaced as the pulse duration shortens to $0.5T_0$. This trend is quantified in Fig.~\ref{fig6}(d), wherein the average peak width is largest for $\tau=0.5T_0$ and decreases as the pulse duration increases to $\tau=1.0T_0$.

To clarify the physical origin of this behavior, we analyze the ionization dynamics of the contributing trajectories, similar to the analysis of FSH stripe density. Figure~\ref{fig7}(a) displays the ionization times of the direct and backward-scattered trajectories as a function of the final parallel momentum, plotted for trajectories with zero final perpendicular momentum for clarity. Similar to the high-order harmonic generation (HHG)~\cite{Lewenstein1994_PRA, Lewenstein1995_PRA}, two distinct backward STs lead to the same final momentum. Depending on their excursion ( recollision) time $\Delta t$, these trajectories are classified as long or short STs~\cite{Li2014_PRA, Shvetsov2018_PRA}. The ionization times of the DTs remain similar across pulse durations, except at high parallel momenta, where differences arise from the varying cutoff energies. In contrast, both the short and long STs exhibit a pronounced dependence on pulse duration, particularly the short STs, whose ionization times vary more strongly for a given $p_{\parallel}$. Figure~\ref{fig7}(b) further shows that the recollision time for both short and long STs is shortest for the $\tau = 0.5T_0$ pulse and increases gradually with longer pulse durations. A larger temporal separation between ionization and rescattering enhances the accumulated phase difference between interfering trajectories, resulting in a denser interference pattern~\cite{Borbely2013_PRA,Yuan2021_PRA}. Consequently, the BSH fringes appear most closely spaced for the $\tau = 1.0T_0$ pulse and become progressively broader as the pulse duration decreases. 

\begin{figure}[t]
\centering\includegraphics[width=1\columnwidth]{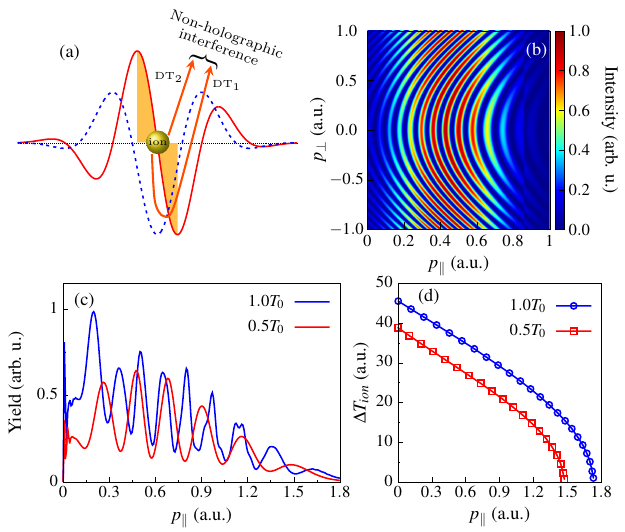}
\caption{Variation in fringe spacing of the non-holographic interference structure. (a) Schematic illustration of the intracycle time double-slit interference trajectories . The half-cycle laser field $\mb{E}(t)$ (red solid curve) with CEP $\phi_0 = 90^\circ$ and the corresponding vector potential $\mb{A}(t)$ (blue dashed curve) are shown. The shaded quarter-cycles indicate the ionization times of the interfering DTs. (b) PMDs showing corresponding interferometric structures by the classical-trajectory based model. (c) Line profiles of the TDSE-simulated PMD spectrum [given in Fig.~\ref{fig3}] for CEP $\phi_0 = 90^\circ$ laser fields, obtained by integrating the spectrum over polar angle $\theta$. (d) Difference of ionization times $\Delta T_{ion}$ of the two interfering DTs, extracted for trajectories with zero final perpendicular momentum using the classical model. }
\label{fig8}
\end{figure}

So far we have analyzed the pulse duration dependence of holographic structures arising from the interference between rescattered and non-rescattered (direct) electrons. We now turn to the non-holographic intracycle interference pattern, which originates from the interference between two non-rescattering trajectories, commonly referred to as the time double-slit (TDS) interference. Figure~\ref{fig8} illustrates how the pulse duration influences the spacing of these interference fringes. In Fig.~\ref{fig8}(a), a schematic diagram depicts the relevant electron trajectories for a half-cycle laser pulse with CEP $\phi_0 = 90^\circ$. The two direct trajectories, DT$_1$ and DT$_2$, are ionized during different quarter-cycles of the field (indicated by the shaded region). Both trajectories avoid rescattering but reach the detector with the same final momentum: one  (DT$_1$) goes across the core without being deflected, while the other (DT$_2$) reaches the detector directly. The time double-slit interference arise from the coherent superposition of these two trajectories, which are born at times $t_{i1}$ and $t_{i2}$ satisfying $A(t_{i1}) = A(t_{i2})$, where the electric field values are of opposite sign, $E(t_{i1}) = -E(t_{i2})$~\cite{Bian2011_PRA,Gopal2009_PRL}. This condition ensures that the two electron wave packets acquire identical final momenta but accumulate different emission phases, giving rise to the characteristic intracycle interference pattern. Fig.~\ref{fig8}(b) shows the intracycle interferometric structures obtained using the classical-trajectory model. The calculated patterns qualitatively reproduce the rightward-curved semiring structures observed in the positive parallel momentum region of TDSE-simulated PMDs presented in Fig.~\ref{fig3}, confirming that the classical framework successfully captures the essential temporal interference mechanism. 

Figure~\ref{fig8}(c) presents the line profiles of the TDSE-simulated PMD spectra [shown in Fig.~\ref{fig3}], obtained by integrating the distributions over the polar angle $\theta$. A clear trend is observed: as the pulse duration decreases from $\tau = 1.0T_0$ to $0.5T_0$, the separation between consecutive peaks in the above-threshold ionization (ATI) spectrum increases. 

For multi-cycle laser pulses, the intracycle modulation is known to be independent of the total number of cycles~\cite{Diego2010_PRA}. In contrast, a pronounced dependence emerges in the sub-cycle regime. This behavior originates from the intrinsic chirp of the sub-cycle pulses, which  effectively induces a blueshift in the driving field frequency, thereby increasing the spacing between adjacent interference fringes. To verify this interpretation, we analyze the ionization dynamics of the interfering trajectories DT$_1$ and DT$_2$ at $p_{\perp} = 0$ a.u. for the two pulse durations. Figure~\ref{fig8}(d) shows the difference in their ionization times, $\Delta T_{ion}$ as a function of the final parallel momentum $p_{\parallel}$. The time difference $\Delta T_{ion}$ decreases with increasing $p_{\parallel}$ for both pulse durations. Moreover, at a fixed $p_{\parallel}$, the time difference $\Delta T_{ion}$ is smaller for the shorter pulse. A smaller time separation between the two ionization events leads to a larger fringe spacing in the time double-slit (intracycle) interference. This observation confirms that the pulse-duration-dependent variation of intracycle interference fringes is governed by the temporal separation of the contributing ionization events, which in turn reflects the influence of the intrinsic chirp of the sub-cycle pulses.

\begin{figure}[t]
\centering\includegraphics[width=1\columnwidth]{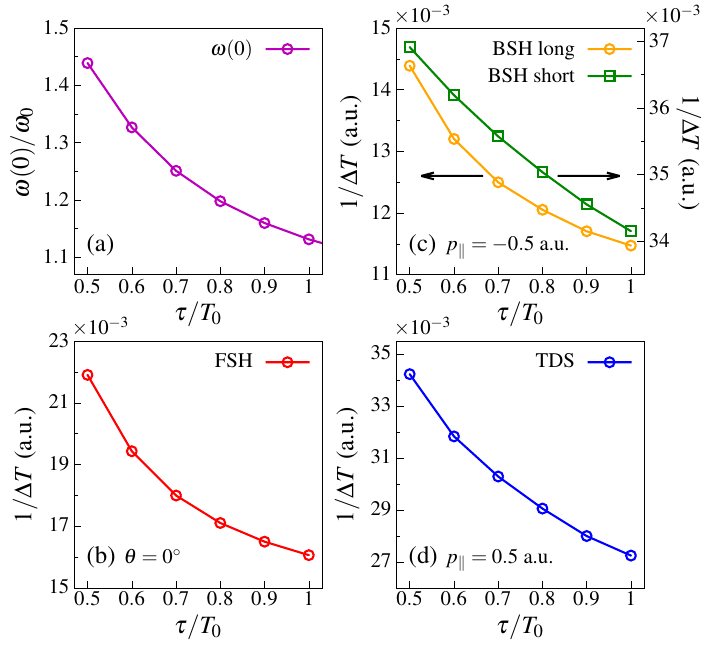}
\caption{(a) Relative blueshift of the instantaneous laser frequency at the center ($t=0$) as a function of pulse durations. Panels (b)-(d) show the dependence of the recollision time $\Delta T$ on pulse duration for the (b) FSH, (c) BSH, and (d) time double-slit (TDS) interference structures. The recollision times $\Delta T$ are extracted at fixed polar angle $\theta$ (FSH) or fixed parallel-momentum $p_{\parallel}$ (BSH and TDS), as indicated in each panel.}
\label{fig9}
\end{figure} 

Finally, we examine the causal link between the intrinsic chirp of sub-cycle pulses and the recollision time $\Delta T$ associated with holographic and non-holographic interference structures. This analysis strengthen the assertion that the intrinsic chirp governs the temporal properties of the interfering trajectories and thereby controls the resulting fringe spacings. Figure~\ref{fig9}(a) shows the relative blueshift of the center frequency $\omega(0)$ as a function of pulse duration. As the pulse length decreases from one optical cycle to the sub-cycle regime, the center frequency $\omega(0)$ exhibits a marked increase, reflecting the enhanced intrinsic chirp. This chirp effectively compresses the temporal waveform, producing an inverse relation between the center frequency and the electron recollision time $\Delta T$.

In Figs.~\ref{fig9}(b)-\ref{fig9}(d), we further illustrate this relationship for FSH, BSH, and TDS interference structures by plotting $1/\Delta T$ as a function of the pulse duration $\tau$. The recollision times $\Delta T$ are extracted at fixed polar angle $\theta$ (FSH) or fixed parallel-momentum $p_{\parallel}$ (BSH and TDS), as indicated in each panel. In all three cases, the dependence of $1/\Delta T$ on $\tau$ mirrors the trend of the center-frequency blueshift, confirming that shorter pulses with stronger intrinsic chirp yield shorter recollision times. This behavior is fully consistent with the variations in the average peak widths (PWs) shown in Figs.~\ref{fig4}(d) and \ref{fig6}(d), which display the same monotonic dependence on pulse duration. Together, these results demonstrate unambiguously that the intrinsic chirp of sub-cycle driving fields modulates the recollision times of the contributing trajectories, and that this modulation directly governs the fringe spacings of the FSH, BSH, and TDS interference patterns.

\subsection{Effect of pulse envelope}
\label{subsec3}
In the preceding subsection, we examined how the number of laser cycles influences the holographic and non-holographic interference structures. When the pulse duration enters the sub-cycle regime, the \textit{intrinsic chirp} of the driving pulse becomes an essential determinant of these structures. We now investigate how the interference patterns evolve when half-cycle pulses with different envelope functions are employed.

\begin{figure}[t]
\centering\includegraphics[width=1\columnwidth]{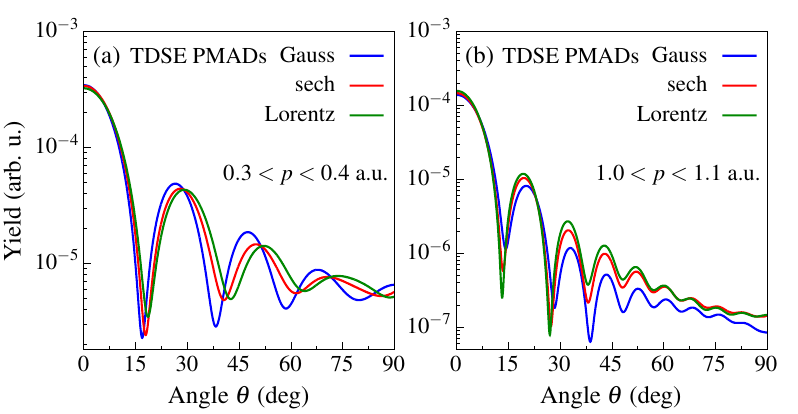}
\caption{Effect of laser pulse envelope on the FSH interferometric structures. Panels (a) and (b) show the variation in stripe density of the spider-leg-like FSH pattern. The PMADs are obtained by integrating the PMDs along the radial direction over a narrow momentum interval: (a) $0.3 < p < 0.4$ a.u. and (b) $1.0 < p < 1.1$ a.u. All results are extracted from TDSE-simulated PMDs of helium driven by half-cycle pulses with various envelope functions [defined in Eq.~\ref{laser_envlp}] and CEP $\phi_0 = 0^\circ$.}
\label{fig10}
\end{figure}

\begin{table}[b]
\caption{\label{tab1} Average peak widths of PMADs presented in Fig.~\ref{fig10} for various pulse envelopes. }
\begin{ruledtabular}
\begin{tabular}{lcc}
Envelope & \multicolumn{2}{c}{Average PWs (deg)}\\
shape    & $0.3<p<0.4$ a.u. & $1.0<p<1.1$ a.u.\\ 
\colrule
Gaussian & 20.04 & 14.06\\
Hyperbolic secant & 21.45 & 13.36\\
Lorentzian & 22.15 & 13.34\\
\end{tabular}
\end{ruledtabular}
\end{table}

Figure~\ref{fig10} illustrates the impact of pulse envelopes on the FSH structures. The PMADs associated with the FSH pattern were extracted from the TDSE-simulated PMDs of helium driven by half-cycle pulses with the envelope functions defined in Eq.~\ref{laser_envlp} and a CEP of $\phi_0 = 0^\circ$. The envelope dependence of the FSH stripe density exhibits two opposite trends depending on the electron momentum. Figures~\ref{fig10}(a) and \ref{fig10}(b) show PMADs obtained by integrating the PMDs over the radial intervals $0.3<p<0.4$ a.u. and $1.0<p<1.1$ a.u., respectively. At lower momenta [Fig.~\ref{fig10}(a)], the FSH stripes are most densely spaced for the Gaussian envelope and become progressively sparser for the sech and Lorentzian envelopes. At higher momenta [Fig.~\ref{fig10}(b)], this trend reverses: the Lorentzian pulse produces the highest stripe density, followed by the sech and then the Gaussian pulses. A quantitative comparison of the average peak widths (PWs) of PMADs for the three pulse envelopes is provided in Table~\ref{tab1}. In the low-momentum interval ($0.3<p<0.4$ a.u.), the Gaussian pulse yields the smallest PW and the Lorentzian pulse the largest, whereas in the high-momentum interval ($1.0<p<1.1$ a.u.), the ordering is reversed. These contrasting behaviors highlights the sensitivity of the FSH interference pattern to the temporal shape of sub-cycle driving fields.

\begin{figure}[t]
\centering\includegraphics[width=1\columnwidth]{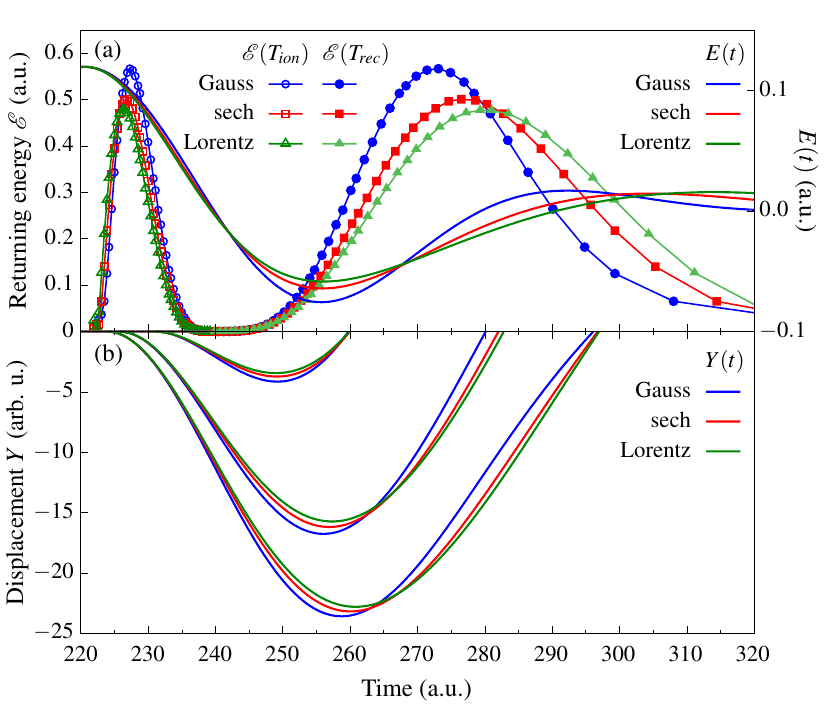}
\caption{Returning energies and classical trajectories of electrons ionized during the quarter-cycle following the peak of half-cycle laser pulses with CEP $\phi_0 = 0^\circ$. (a) Returning energies ($\mathcal{E}$) at the ionization times ($T_{ion}$) and rescattering times ($T_{res}$), along with the corresponding laser electric fields (simple lines, right-axis) for different pulse envelopes. The full waveforms of half-cycle fields are presented in Fig.~\ref{fig_laser_wt}(a). (b) Classical electron trajectories initiated at three representative ionization times within the first quarter-cycle after the pulse maximum, shown for laser pulses with different envelope functions.}
\label{fig11}
\end{figure}

To elucidate the origin of these trends, we examine the returning energies and classical trajectories of electrons ionized during the quarter-cycle following the field maximum, shown in Fig.~\ref{fig11}. The returning energies $\mathcal{E}$ are largest for the Gaussian pulse and decreases systematically for the sech and Lorentzian pulses. This arises because the Gaussian pulse have a larger field amplitude in the subsequent half-cycle after ionization. This stronger post-ionization field accelerates electrons to higher return energies and yields larger excursion distances [Fig.~\ref{fig11}(b)]. Moreover, for a fixed returning energy, the recollision time $\Delta T$ is shortest for the Gaussian pulse and increases for the sech and Lorentzian pulses.

The direct and rescattered trajectories respond differently to the envelope. Because all envelopes have similar fields during the quarter-cycle that initiates the trajectories, the phase accumulated by the direct electrons (determined mainly by the vector potential near ionization), varies only weakly among the three pulses. The scattering trajectories, however, are strongly influenced by the amplitude of the subsequent half-cycle. The Gaussian envelope, having the largest field amplitude in this region, drives electrons to larger excursions and higher return energies, producing a larger phase accumulation along the rescattered path. This increases the phase difference between the direct and scattering trajectories and therefore reduces the fringe spacing, yielding a higher stripe density in the FSH pattern for the Gaussian pulse.

A competing mechanism arises from the intrinsic chirp of the sub-cycle pulses. As shown in Fig.~\ref{fig_laser_wt}, the Gaussian field exhibits a stronger temporal compression of its wings due to its larger instantaneous-frequency variation. This reduces the recollision time $\Delta T$ and tends to widen the FSH fringes, leading to a lower stripe density. This behavior parallels the trend observed in the cycle-number analysis of Fig.~\ref{fig4}.

The envelope-dependent stripe density thus results from the competition between two mechanisms. On the one hand, the stronger post-ionization field amplitude of the Gaussian pulse leads to greater phase accumulation along the scattering trajectory and consequently denser FSH fringes. On the other hand, the larger instantaneous-frequency of the Gaussian field shortens the recollision time $\Delta T$, which tends to widen the holographic fringes and reduce their density. The observed stripe density trend in the PMADs is determined by the relative strength of these two effects in different momentum regions. For lower momenta ($0.3<p<0.4$ a.u.), the amplitude-driven excursion mechanism dominates, resulting in decreasing stripe density in the order of Gaussian to sech to Lorentzian. In contrast, at higher momenta ($1.0<p<1.1$ a.u.), the chirp-induced reduction of $\Delta T$ becomes the dominant factor, reversing the trend so that Lorentzian pulse yields the most densely spaced FSH stripes, followed by the sech and Gaussian envelopes.

\begin{figure}[t]
\centering\includegraphics[width=1\columnwidth]{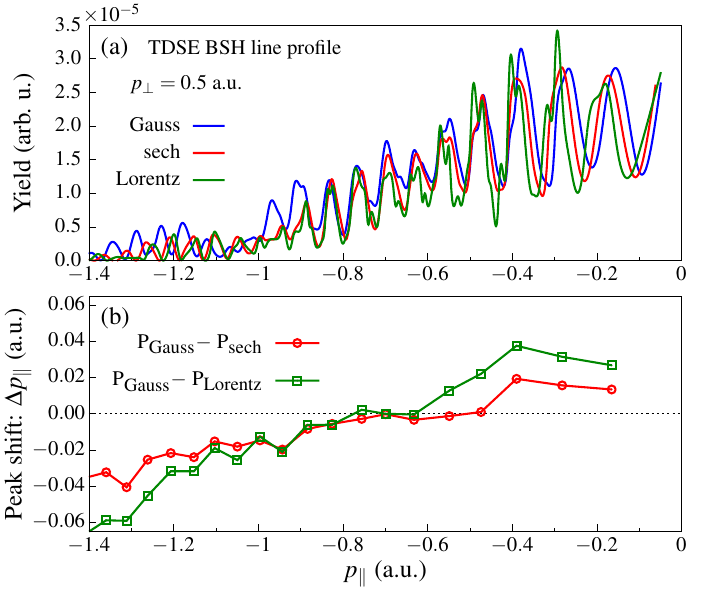}
\caption{Influence of the pulse envelope on the BSH interference pattern. (a) Line profiles of the BSH structure at $p_{\perp} = 0.5$ a.u. for different pulse envelopes defined in Eq.~\ref{laser_envlp}. The profiles are extracted from TDSE-simulated PMDs of helium driven by half-cycle pulses with CEP $\phi_0 = 0^\circ$. (b) Relative shift of the BSH peak positions for the sech (red) and Lorentzian (green) envelopes with respect to those obtained using the Gaussian envelope. }
\label{fig12}
\end{figure}

We now turn to the influence of the pulse envelope on the spacing of the BSH interference peaks. Figure~\ref{fig12}(a) shows the line profiles of the BSH structure at $p_{\perp}=0.5$ a.u. for the envelope functions defined in Eq.~\ref{laser_envlp}. The line-outs were extracted from TDSE-simulated PMDs of helium driven by half-cycle pulses with CEP $\phi_0 = 0^\circ$. Near the origin ($p_{\parallel}=0$), the fringe spacing is largest for the Lorentzian envelope and successively decreases for the sech and Gaussian pulses. As $p_{\parallel}$ becomes more negative, the separation between adjacent peaks decreases for all envelopes, and the ordering eventually reverses near the cutoff: the Gaussian pulse produces the widest spacing, followed by the sech pulse, while the Lorentzian pulse yields the narrowest. This trend is quantified in Fig.~\ref{fig12}(b), which shows the relative shift of the BSH peak positions for the sech and Lorentzian envelopes with respect to those obtained using the Gaussian envelope. Both envelopes exhibit positive peak shift near the origin, indicating larger fringe spacing then that of the Gaussian pulse; the shift decreases toward intermediate momenta, crosses zero, and becomes negative in the cutoff region, demonstrating that the Gaussian-pulse fringes become more widely separated at higher $|p_{\parallel}|$. This observed behavior reflects the same competition identified for the FSH structure. Close to the origin, the dominant contribution arises from the stronger post-ionization field amplitude of the Gaussian pulse, which drives the rescattered electron to larger excursion distances and thus yields greater phase accumulation, producing more closely spaced fringes. At larger $|p_{\parallel}|$, however, the intrinsic chirp of the Gaussian pulse becomes the controlling factor: the higher instantaneous frequency shortens the recollision time $\Delta T$, reducing the differential phase between the interfering trajectories and thereby increasing the fringe spacing.

\begin{figure}[t]
\centering\includegraphics[width=1\columnwidth]{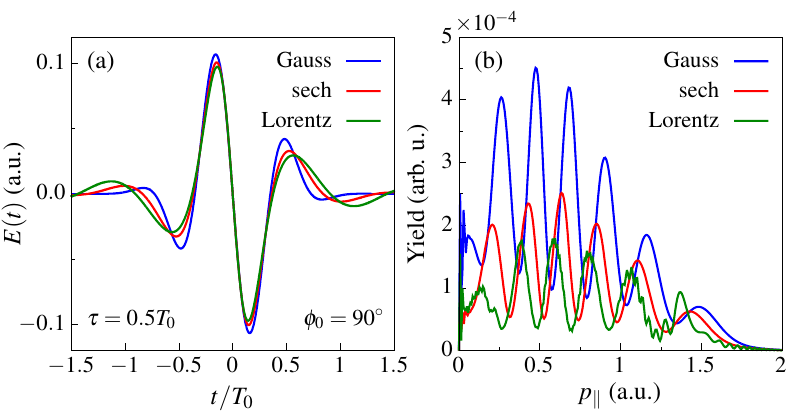}
\caption{Effect of the pulse envelope on the intracycle interference structure. (a) Temporal profiles of half-cycle laser fields with CEP $\phi_0 = 90^\circ$ for different envelope functions. (b) Corresponding line profiles of the TDSE-simulated PMDs, obtained by integrating the distribution over the polar angle $\theta$ in the positive parallel-momentum region.}
\label{fig13}
\end{figure}

Finally, we examine how the pulse envelope influences the intracycle interference structure. Figure~\ref{fig13}(a) shows the temporal profiles of half-cycle laser fields with CEP $\phi_0 = 90^\circ$ for the three envelope functions, and Fig.~\ref{fig13}(b) presents the corresponding ATI spectra obtained by integrating the TDSE-simulated PMDs over the polar angle $\theta$ in the positive $p_{\parallel}$ region. A clear envelope-dependent shift of the intracycle fringes is observed: the ATI peaks move outward as the envelope is varied from Lorentzian to sech to Gaussian. This behavior reflects the interference of two direct (non-rescattering) trajectories, born in the quarter-cycles immediately before and after the pulse center. The field amplitude in these two quarter-cycles is largest for the Gaussian envelope and decreases for the sech and Lorentzian pulses. The stronger instantaneous field of the Gaussian pulse imparts a larger quiver energy to the electron, increasing the final kinetic energy of the interfering trajectories. Since the fringe spacing of time double-slit (intracycle) interference grows monotonically with photoelectron energy~\cite{Diego2006_PRA,Diego2010_PRA}, the higher quiver energy associated with the Gaussian driver shifts the intracycle fringes outward relative to the sech and Lorentzian cases.

Overall, the pulse envelope plays a decisive role in shaping all three types of interferometric structures in sub-cycle regime. For the FSH and BSH patterns, the envelope dependence arises from the competition between two mechanisms: (i) the post-ionization field amplitude, which governs the excursion, return energy, and phase accumulation of the scattering trajectory, and (ii) the intrinsic chirp of the sub-cycle waveform, which controls the recollision time and therefore the fringe spacing. The relative strength of these competing effects varies across momentum regions, leading to envelope-dependent reversals in FSH stripe density and BSH fringe spacing. In contrast, the time double-slit interference is primarily controlled by the instantaneous field amplitude in the half-cycle across the pulse center, which sets the quiver energy of the non-rescattering trajectories and thereby shifts the ATI fringes. Taken together, these results demonstrate that the envelope of a sub-cycle pulse is not merely a multiplicative shaping function but an active control parameter that reshapes strong-field interferometric structures by modifying both the electron dynamics and the underlying temporal phase relationships.

We note that the 3D-TDSE employed in this work rely on the SAE approximation, which neglects electron-electron correlation and may therefore introduce quantitative deviations in regimes involving large-angle rescattering, such as BSH contribution. Nonetheless, the SAE framework has been shown to capture the essential phase dynamics and interference mechanisms with high reliability in strong near-infrared fields~\cite{Brennecke2020_PRL, Khurelbaatar2024_LightSci, Yu2020_PRA}. Since our conclusions are based on relative trends, specifically the dependence of fringe spacing and visibility on pulse duration, intrinsic chirp, and envelope shape. The SAE model is sufficient for the present qualitative analysis, while a fully correlated treatment would be required for quantitatively exact predictions.

\section{Summary and Concluding Remarks}
\label{sec4}

We have systematically studied the effects of sub-cycle driving fields on the interference structures in PMDs, based on the numerical solutions of the TDSE and the classical model. Our results reveal that photoelectron interference patterns generated by sub-cycle laser pulses are highly sensitive to both the pulse duration and the temporal waveform imposed by the pulse envelope. For single- and half-cycle Gaussian pulses with CEP values of $0^\circ$ and $90^\circ$, the TDSE simulations clearly identify three principal contributions to the PMDs: forward-scattering holographic (FSH) interference, backscattering holographic (BSH) interference, and intracycle (time double-slit) interference structures. The visibility and relative dominance of these structures depend strongly on the temporal symmetry of the electric field and vector potential.

For CEP $\phi_0=0^\circ$, half-cycle pulses isolate the BSH fringes, enabling a clean observation of backward-scattered trajectories that are typically obscured in longer pulses. As the pulse duration increases to one optical cycle, intracycle interference becomes prominent and increasingly overlaps with the holographic contributions, complicating the PMD structure. For CEP $\phi_0 = 90^\circ$, the temporally antisymmetric field promotes time double-slit interference, producing spider-leg fringes and fish-bone-like structures whose spacing increases with photoelectron energy.

A detailed analysis of the pulse-length dependence demonstrates that the fringe densities of the FSH, BSH, and intracycle patterns are all governed by the temporal dynamics of the underlying electron trajectories. Classical trajectory analysis shows that the stronger intrinsic chirp exhibited by shorter pulses, modifies the ionization and rescattering times of interfering trajectories. For both FSH and BSH, shorter pulses yield smaller recollision time $\Delta T$, producing more widely spaced interference fringes. The intracycle interference fringes, unlike multi-cycle pulses, exhibit a pronounce dependence on the cycle number of sub-cycle driving fields. In the sub-cycle regime, intrinsic chirp reduces the temporal separation between the two interfering trajectories, resulting in increased ATI peak spacing. Across all three mechanisms, the intrinsic chirp emerges as the unifying factor controlling the relative phase accumulation between interfering electron pathways.

Finally, we examined how the pulse envelope affects the interference patterns for half-cycle pulses. Despite having identical FWHM duration, different envelope functions modify the effective instantaneous frequency and the temporal slope of the vector potential near the ionization times. As a result, both the holographic and intracycle interference fringe spacings exhibit noticeable variations depending on the choice of envelope, further highlighting the sensitivity of strong-field holography to sub-cycle temporal shaping.

In conclusion, our analysis shows that sub-cycle pulses provide a powerful means of tailoring electronic wave packet interference in strong-field ionization. The controllable modification of ionization and rescattering times---via CEP, pulse duration, and pulse envelope---offers a route to selectively enhance or suppress specific holographic features. These findings provide essential guidance for using sub-cycle fields in ultrafast imaging and for exploiting holographic structures as probes of sub-angstrom scale attosecond electronic dynamics.

\section*{Acknowledgments} Authors would like to acknowledge the financial support from the Ministry of Education, Culture, Sports, Science and Technology of Japan (MEXT) through Grants-in-Aid under grant no. 21H01850, and the MEXT Quantum Leap Flagship Program (Q-LEAP) (grant no. JP-MXS0118068681). This project was supported by the RIKEN TRIP initiative (Leading-edge semiconductor technology).

\bibliography{Bibliography}

\end{document}